\documentclass[
prc,%
10pt,%
final,%
notitlepage,%
oneside,%
twocolumn,%
nobibnotes,%
nofootinbib,% In the current version of REVTeX, when this option is on,
superscriptaddress,%
floatfix,%
floatfix,%
showkeys,%
showpacs]%
{revtex4}
\usepackage{color}
\usepackage{amsfonts}
\usepackage{amsbsy}
\usepackage{mathrsfs}
\usepackage{graphicx}
\def\lsim{\mathrel{\rlap{
\lower4pt\hbox{\hskip-3pt$\sim$}}
    \raise1pt\hbox{$<$}}}     %less than approx. symbol
\def\gsim{\mathrel{\rlap{
\lower4pt\hbox{\hskip-3pt$\sim$}}
    \raise1pt\hbox{$>$}}}     %greater than or approx. symbol
\def\scr#1{\mbox{\scriptsize #1}}
\begin{document}
\title{
Bulk Properties of the Matter 
Produced 
%in Heavy-Ion Collisions 
at Energies of 
the Beam Energy Scan Program} 
\author{Yu. B. Ivanov}\thanks{e-mail: Y.Ivanov@gsi.de}
\affiliation{National Research Centre "Kurchatov Institute", Moscow 123182, Russia} 
\affiliation{National Research Nuclear University "MEPhI",  
%(Moscow Engineering Physics Institute),
%Kashirskoe sh. 31, 
Moscow 115409, Russia}
\affiliation{Bogoliubov Laboratory of Theoretical Physics, JINR, Dubna 141980, Russia}
\author{A. A. Soldatov}\thanks{e-mail: saa@ru.net}
\affiliation{National Research Nuclear University "MEPhI",  
%(Moscow Engineering Physics Institute),
%Kashirskoe sh. 31, 
Moscow 115409, Russia}
\begin{abstract}

Recent STAR data on the bulk observables in the energy range of 
the Beam-Energy Scan Program at the Relativistic Heavy-Ion Collider 
are analyzed within the model of the three-fluid dynamics (3FD).
The simulations  are performed with different equations of state (EoS). 
The purely hadronic EoS fails to reproduce the data. 
A good, though imperfect, overall reproduction of the data is found within the deconfinement scenarios. 
The crossover EoS turns out to be slightly preferable. 
For this reproduction a fairly strong baryon stopping in the quark-gluon phase is required.
% which results in rather high initially equilibrated baryon densities in the central region.
The 3FD model does not need two separate freeze-outs 
(i.e. kinetic and chemical ones) to describe 
the STAR data. A unified freeze-out is applied at all energies. 
\pacs{25.75.-q,  25.75.Nq,  24.10.Nz}
\keywords{relativistic heavy-ion collisions, hydrodynamics, bulk observables}
\end{abstract}
\maketitle
%\today

\section{Introduction}

Searching for the critical point and
phase boundary in the QCD phase diagram is 
the main motivation for the Beam-Energy Scan (BES) Program 
at the Relativistic Heavy-Ion Collider (RHIC). 
Recently, experimental results from the systematic study of bulk properties
of the matter produced in Au+Au collisions at BES RHIC energies were published 
by the STAR collaboration \cite{Adamczyk:2017iwn}.
These bulk properties can hardly indicate the presence or absence of the 
critical point, because the latter are expected to be manifested in 
enhanced fluctuations in multiplicity
distributions of conserved quantities \cite{Stephanov:2008qz,Karsch:2010ck}. 
The systematic study of these bulk properties
may reveal the evolution and change in behavior of the
system formed in heavy-ion collisions as a function of
collision energy.
In particular, these bulk observables may indicate if the 
quark confinement
and chiral symmetry breaking at finite baryon densities
occur simultaneously or they are separated in the phase diagram 
as argued in Ref. \cite{Suganuma:2017syi}.

In the present paper the bulk observables in the BES-RHIC energy range
are analyzed within the model of three-fluid dynamics
(3FD) \cite{3FD}. 
The 3FD model is quite successful in reproduction 
of the major part of bulk
observables at energies of the Alternating Gradient
Synchrotron (AGS) at BNL and the  
Super Proton Synchrotron (SPS) at CERN: 
the baryon stopping \cite{Ivanov:2013wha,Ivanov:2012bh}, 
yields of different hadrons, their rapidity \cite{Ivanov:2013yqa}
and transverse momentum \cite{Ivanov:2013yla}
distributions. 
%of the European Organization for Nuclear Research (CERN).  
It is also applicable to the description of the elliptic \cite{Ivanov:2014zqa} 
and directed \cite{Konchakovski:2014gda,Ivanov:2016sqy} flow  
in the AGS-SPS-BES/RHIC energy range. 
Predictions of the 3FD model for some of the bulk observables at BES-RHIC energies 
have been already reported in Refs. 
\cite{Ivanov:2012bh,Ivanov:2013yla,Ivanov:2016sqy} and fragmentarily 
compared with preliminary STAR data. 
The present paper is devoted to the systematic analysis of the final STAR data on the bulk observables.

%\cite{Ivanov:2013wha,Ivanov:2012bh,Ivanov:2013yqa,Ivanov:2013yla,Ivanov:2014zqa,Konchakovski:2014gda,Ivanov:2016sqy} 

\section{3FD model}
\label{Model}

%Unlike the conventional hydrodynamics, where local
%instantaneous stopping of projectile and target matter is
%assumed, a specific feature of the 3FD description \cite{3FD} 
%is a finite stopping power resulting in a counterstreaming
%regime of leading baryon-rich matter. 

A specific feature of high-energy heavy-ion collisions is a finite stopping power 
resulting in a counterstreaming regime of baryon-rich matter of the projectile and target 
rather then their instantaneous stopping. % of projectile and target matter. 
Within the 3FD description \cite{3FD}
this generally
nonequilibrium regime of the baryon-rich matter
is modeled by two interpenetrating baryon-rich fluids 
initially associated with constituent nucleons of the projectile
(p) and target (t) nuclei. In addition, newly produced particles, predominantly
populating the midrapidity region, are associated with a fireball
(f) fluid.
Each of these fluids is governed by conventional hydrodynamic equations 
coupled by friction terms in the right-hand sides of the Euler equations. 
The hydrodynamic equations do not include viscosity because the dissipation is 
provided by the friction terms. 
These friction terms describe energy--momentum loss of the baryon-rich fluids. 
A part of this
loss is transformed into thermal excitation of these fluids, while another part 
gives rise to particle production into the fireball fluid.

Friction forces between fluids are the key constituents of the model that 
determine dynamics of the nuclear collision. 
The friction forces in the hadronic phase were estimated in Ref. \cite{Sat90}. 
%Precisely these friction forces are used in the simulations 
%for the hadronic phase. 
There are no theoretical estimates of
the friction in the quark-gluon phase (QGP) so far.
Therefore, the friction in the QGP is purely phenomenological.
It was fitted to reproduce the baryon
stopping at high incident energies within the deconfinement
scenarios as described in  Ref. \cite{Ivanov:2013wha} in detail.

The physical input of the present 3FD calculations is described in
Ref.~\cite{Ivanov:2013wha}. 
The simulations in 
\cite{Ivanov:2013wha,Ivanov:2012bh,Ivanov:2013yqa,Ivanov:2013yla,Ivanov:2014zqa,Konchakovski:2014gda,Ivanov:2016sqy} 
were performed with different 
equations of state (EoS's)---a purely hadronic EoS \cite{gasEOS}  
and two versions of the EoS involving the   deconfinement
 transition \cite{Toneev06}, i.e. a first-order phase transition  (1st-order-tr.)
and a smooth crossover one. 
It was demonstrated that the purely hadronic EoS fails to reproduce 
data at high collision energies ($\sqrt{s_{NN}}\gsim$ 5 GeV). 
Therefore, in the present paper we demonstrate results with 
only these deconfinement
%first-order-transition and crossover 
EoS's as the most successful in reproduction of various 
bulk observables at high collision energies and only occasionally 
present results of the purely hadronic EoS.   

\begin{figure}[htb]
%\vspace*{-14mm}
\includegraphics[width=7.2cm]{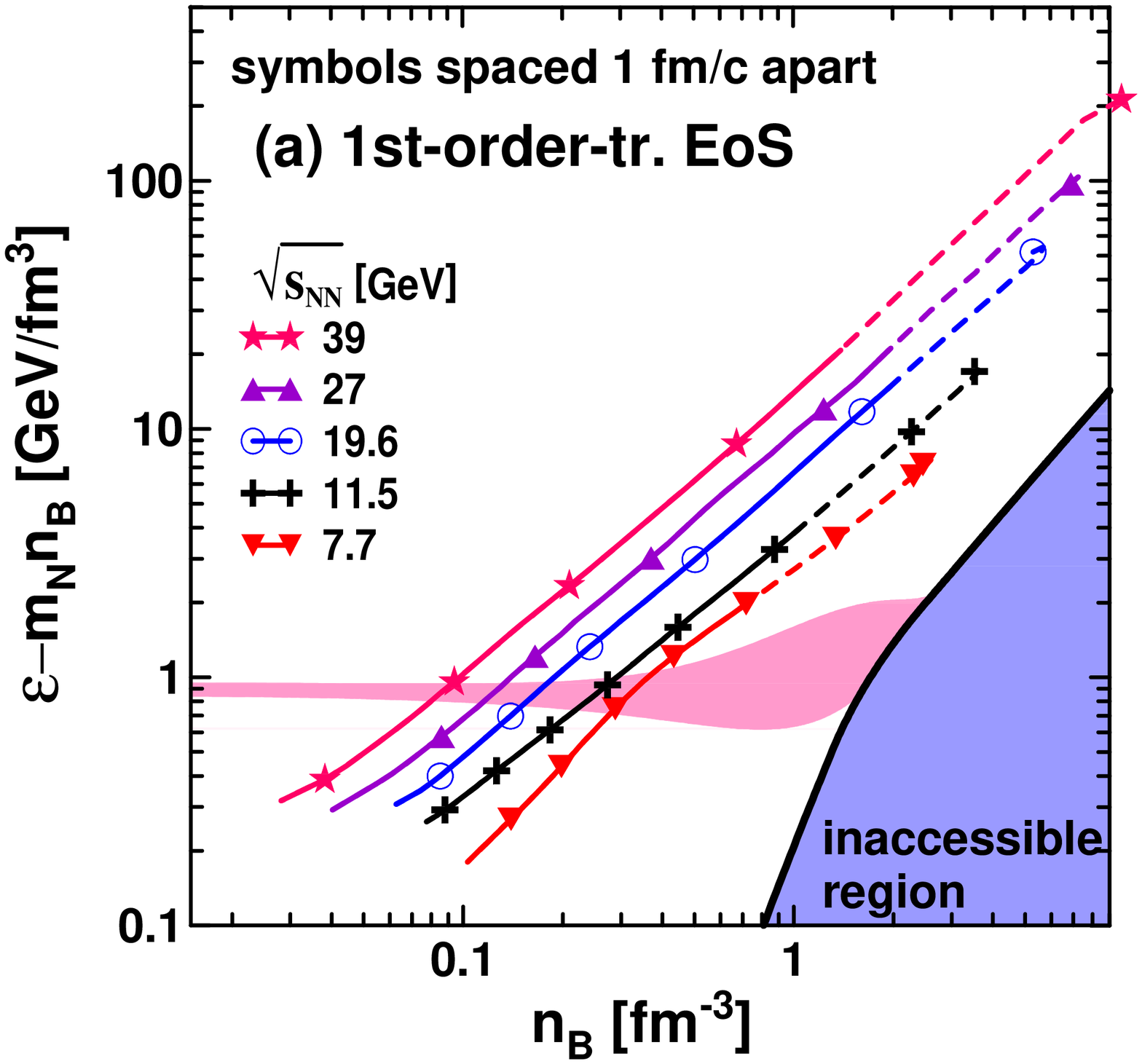}
\\
\includegraphics[width=7.2cm]{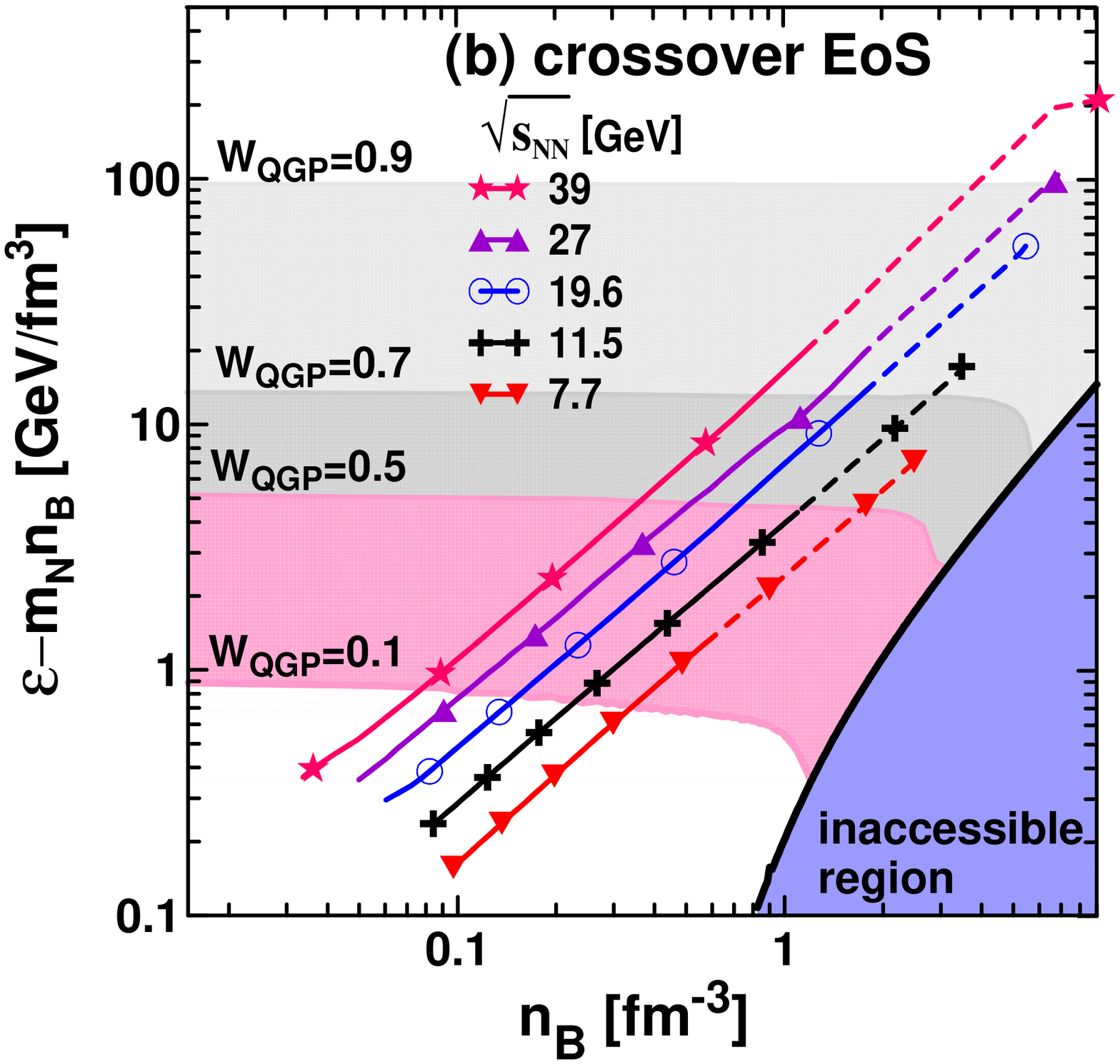}
 \caption{(Color online)
Dynamical trajectories of the matter in the central region of 
colliding Au+Au nuclei  
%(4fm$\times$4fm$\times \gamma_{cm}$4fm), where $\gamma_{cm}$ is the Lorentz
%factor associated with the initial nuclear motion in the c.m. frame, 
%in central collisions 
($b=$ 2 fm) at BES-RHIC energies. 
%$\sqrt{s_{NN}}=$ 7.7, 11.5, 19.6, 27 and 39 GeV. 
The trajectories are plotted in terms
of the proper (i.e. in the local rest frame) baryon density ($n_B$) and 
the proper energy density minus $n_B m_N$ ($\varepsilon - m_N n_B$)
where $m_N$ is the nucleon mass. 
Only expansion stages of the
evolution are displayed.  
Symbols on the trajectories indicate the time rate of the evolution:
time span between marks is 1 fm/c. 
Evolution of the equilibrated matter by solid lines, while 
the stage before the equilibration is displayed by dashed lines. 
The trajectories are presented 
for two EoS's: (a) the first-order-transition (1st-order-tr.) EoS and (b) the crossover EoS. 
For the 1st-order-tr. EoS the mixed phase is displayed by 
the shadowed region. 
% is located between the borders, 
%where the QGP phase start to raise ($W_{QGP}=$ 0) and becomes completely 
%formed ($W_{QGP}=$ 1). 
For the crossover EoS   the displayed borders correspond to
values of the QGP fraction $W_{QGP}=$ 0.1, 0.5, 0.7 and 0.9.
The inaccessible region is restricted by $\varepsilon(n_B,T=0)-m_N n_B$ from above. 
}
\label{fig1}
\end{figure}

Figure \ref{fig1} illustrates the dynamics of nuclear collisions at BES-RHIC energies.  
Similarly to Ref. \cite{Randrup07}, 
the figure displays dynamical 
trajectories of the matter in the central box placed around the
origin ${\bf r}=(0,0,0)$ in the frame of equal velocities of
colliding nuclei:  $|x|\leq$ 2 fm,  $|y|\leq$ 2 fm and $|z|\leq$
$2 \gamma_{cm}$ fm, where $\gamma_{cm}$ is the Lorentz
factor associated with the initial nuclear motion in the c.m. frame.  
Initially, the colliding nuclei are placed symmetrically with respect
to the origin ${\bf r}=(0,0,0)$; $z$ is the direction of the beam.
The size of the box was chosen 
to be large enough that the amount of matter in it can be
representative to conclude on the onset of deconfinement 
and to be small enough to consider the matter in it as a homogeneous
medium. Nevertheless, the matter in the box still amounts to a minor part
of the total matter of colliding nuclei.  

Only expansion stages of the evolution are displayed. 
Evolution proceeds from the top point of the trajectory downwards.
Symbols mark the time intervals along the trajectory. 
Subtraction of the $m_N n_B$ term from the energy density is done  for the sake of suitable 
representation of the plot. 
The $\varepsilon$-$n_B$ representation
is chosen because these densities 
are dynamical quantities and, therefore, are suitable to compare
calculations with different EoS's. 
At a given density $n_B$, the zero-temperature
compressional energy, $\varepsilon(n_B,T=0)$, provides a lower bound on
the energy density $\varepsilon$, so the accessible region is correspondingly
limited.

The non-equilibrium stage of the expansion is displayed by dashed lines 
in Fig. \ref{fig1}. 
The criterion of the thermalization is equality of longitudinal  
($P_{\rm{long}}=T_{zz}$)
and transverse 
($P_{\rm{tr}}=[T_{xx}+T_{yy}]/2$)
pressures\footnote{
Note that the spatial components of the collective four-velocity of the composed matter 
in the considered central box are zero due to symmetry reasons. 
}
in the box with accuracy better than 10\%. 
Here 
\begin{eqnarray}
\label{T_tot}
%T^{\mu\nu}_{\scr{tot}} \equiv
T^{\mu\nu} \equiv
T^{\mu\nu}_{\scr p} + T^{\mu\nu}_{\scr t} + T^{\mu\nu}_{\scr f}
\end{eqnarray}
is the energy--momentum tensor of the composed matter
being the sum of conventional hydrodynamical energy--momentum tensors of separate fluids.
In similar figures of Ref.~\cite{Ivanov:2013wha}, the onset of equilibration was 
indicated at earlier stages. The reason is that the condition  
$|P_{\rm{long}}-P_{\rm{tr}}|\leq 0.05 (P_{\rm{long}}+P_{\rm{tr}})$ indeed is first 
reached at earlier times as compared with those indicated in Fig. \ref{fig1}. 
However, after that the difference between $P_{\rm{long}}$ and $P_{\rm{tr}}$
remains at the 10\% level for some time or even slightly exceeds this 10\% level. 
Therefore, in the present Fig. \ref{fig1} we mark the thermalization instant as 
that after which the difference between $P_{\rm{long}}$ and $P_{\rm{tr}}$ 
rapidly drops below the 10\% level.

The trajectories for the first-order-transition  and crossover  
EoS's are very similar, as seen from Fig. \ref{fig1}. 
Note that this similarity is not due to
similarity of these two EoS's. This similarity takes place
because of the friction forces in the QGP phase 
that were independently fitted 
for each EoS \cite{Ivanov:2013wha} in order to reproduce observables in the
midrapidity region.

As seen in Fig. \ref{fig1}, the equilibration in the central region is 
achieved at rather high baryon densities even at 
high collision energies. This in contrast to conclusions made in Refs. 
\cite{Bialas:2016epd,Shen:2017bsr} where formation of the initial state for the 
hydrodynamic evolution was studied. E.g., in Ref. \cite{Bialas:2016epd} it was  
even concluded that the stopped
nucleons from the target and the projectile end up separated from each other by
the distance increasing with the collision energy. For $\sqrt{s_{NN}}>$
 6 or 10 GeV  the latter implies 
that the created system is not in thermal and chemical equilibrium, and the
net-baryon density reached is likely not much higher than that already present in
the colliding nuclei \cite{Bialas:2016epd}.

The equilibration in the central region in collisions at energies up to the top SPS one 
was also analyzed in \cite{Bravina:2008ra} within the Quark-Gluon String Model (QGSM)
and the model of the Ultrarelativistic Quantum Molecular Dynamics (UrQMD). 
The equilibration time in a small box (0.5 fm $\times$ 0.5 fm $\times$ 0.5 fm) 
within the QGSM and UrQMD \cite{Bravina:2008ra} is very similar to the 3FD time
at the energy of 7.7 GeV (or 30A GeV in the lab. frame). 
We do not refer to the results of Ref. \cite{Bravina:2008ra} in the large box 
(5 fm $\times$ 5 fm $\times$ 5 fm) because this box is too large to consider the 
matter in it as a homogeneous medium in view of the Lorentz contraction of colliding 
nuclei. However, at the top SPS energy the QGSM and UrQMD equilibration times are essentially longer 
than that within the 3FD at the similar energy of 19.6 GeV. 
Consequently, the higher equilibrium densities are reached within the 3FD simulations. 
This is the effect of the stronger QGP friction 
required for reproduction of the SPS data 
\cite{Ivanov:2013wha,Ivanov:2012bh,Ivanov:2013yqa,Ivanov:2013yla}. 
Note that even within the hadronic scenario we had to considerably enhance the hadronic friction 
\cite{3FD,Ivanov:2013wha} as compared with its microscopic estimate  \cite{Sat90}
in order to reproduce data at the top SPS energy.

In spite of the rather high initial baryon densities in the center region 
even at high collision energies, cf. Fig. \ref{fig1}, 
the resulting proton rapidity distributions manifest a dip at the midrapidity \cite{Ivanov:2013wha}. 
In fact, there are two ways to achieve the midrapidity dip in the net-baryon rapidity 
distribution. The first one is high transparency of colliding nuclei as implemented 
in Refs. \cite{Bialas:2016epd,Shen:2017bsr}. An alternative scenario is the purely 
hydrodynamical 1D expansion of the initially compressed and thermalized thin disk of the baryon-rich matter. 
In the process of this one-dimensional (1D) expansion the baryon charge is pushed out to peripheral rapidities 
thus producing a dip at the midrapidity. These two scenarios were tested within the 3FD model when 
the friction parameters were fitted \cite{Ivanov:2013wha} at the BES-RHIC energies. 
It was found that the problem of the first ``transparent'' scenario is that too many pions are produced. 
The actual friction of the 3FD model does not result in an immediate stopping of colliding nuclei, 
as seen from Fig. \ref{fig1}. E.g., in the initial thermalized state 
at the collision energy of 39 GeV,  
approximately 30\% of the baryon charge is located in ``true''
fragmentation regions, i.e. in the baryon-rich matter passed through the
interaction region, while 70\% is located in the central fireball \cite{Ivanov:2017xee}. 
Probably, the ``transparent'' scenario is relevant at the top RHIC and LHC energies.

\begin{figure*}[htb]
%\vspace*{-14mm}
\includegraphics[width=18.0cm]{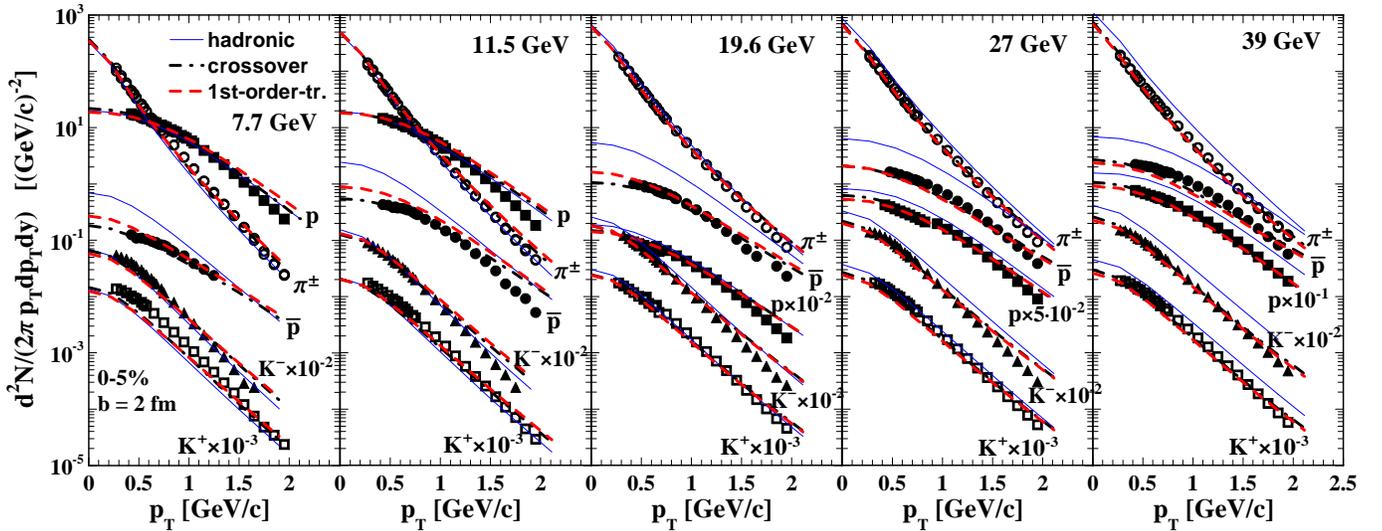}
 \caption{(Color online)
Transverse-momentum spectra at the midrapidity of various particles 
produced in central ($b=2$ fm) Au+Au collisions at $\sqrt{s_{NN}}=$ 7.7 -- 39 GeV predicted
 by the 3FD model with three considered EoS's. 
Experimental STAR data (0-5\% centrality) are from Ref. \cite{Adamczyk:2017iwn}.
}
\label{fig2}
\end{figure*}

\section{Bulk observables}
\label{Bulk observables}

Simulations of Au+Au collisions at energies $\sqrt{s_{NN}}=$ 7.7, 11.5, 19.6, 27 and 39 GeV 
were performed at fixed impact parameters ($b$). The correspondence  between experimental
centrality and the mean impact parameter was taken from
Glauber simulations of Ref. \cite{Abelev:2008ab}. 
Though Ref. \cite{Abelev:2008ab} deals with higher collision energies, 
we implemented the reported correspondence to the BES-RHIC energy range 
in view of practical energy independence of the results deduced in Ref. \cite{Abelev:2008ab}. 
If it is not stated otherwise, proton and antiproton observables include contributions from weak decays, 
%of hyperons, 
while pion and kaon observables do not. This is in agreement with the 
STAR measurements \cite{Adamczyk:2017iwn}.

\subsection{Transverse-Momentum Spectra}
\label{Transverse-Momentum Spectra}

Figure \ref{fig2} demonstrates transverse-momentum spectra at the midrapidity of various particles 
produced in central ($b=2$ fm) Au+Au collisions at $\sqrt{s_{NN}}=$ 7.7 -- 39 GeV predicted
 by the 3FD model with three considered EoS's. 
Experimental STAR data (0-5\% centrality) are from Ref. \cite{Adamczyk:2017iwn}.
As seen, the hadronic scenario fails to describe the data. It fails 
first only for antiprotons at lower energies and then for all hadrons at 
higher collision energies.

The first-order-transition and crossover scenarios 
are quite successful in the reproduction of the data \cite{Adamczyk:2017iwn}
and give very similar results at all considered energies. 
However, there are exceptions. 
The $K^+$ spectra are underestimated at 7.7 GeV and less spectacular at 11.5 GeV. 
This is related to the failure to reproduce the horn in the $K^+/\pi^+$ ratio, discussed below.

\begin{figure}[htb]
%\vspace*{-14mm}
\includegraphics[width=6.2cm]{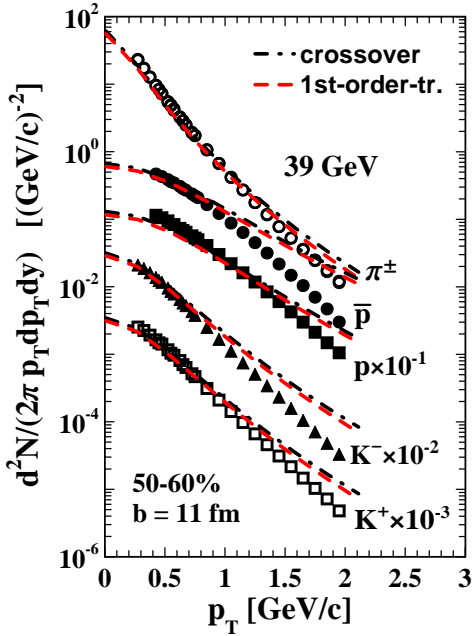}
 \caption{(Color online)
Transverse-momentum spectra at midrapidity of various particles 
produced in non-central ($b=11$ fm) Au+Au collisions at $\sqrt{s_{NN}}=$ 39 GeV predicted
 by the 3FD model with two considered  deconfinement EoS's. 
Experimental STAR data (50-60\% centrality) are from Ref. \cite{Adamczyk:2017iwn}.
}
\label{fig3}
\end{figure}

A common feature of the 3FD results is overestimation of high-$p_T$ tails of the $p_T$ spectra. 
This overestimation happens earlier (i.e. at lower $p_T$) for rare probes ($\bar{p}$ and $K^-$). 
This is a consequence of the 3FD description being based on grand-canonical statistics
which requires large multiplicities to be valid. A lack of
 exact conservations (of the baryon number and strangeness) in
 the grand-canonical ensemble results in overestimation of
 data for rare probes.
Even abundant hadronic probes become
 rare at high momenta. Therefore, their treatment on the basis of
 the grand-canonical ensemble results in overestimation of their
 yields. Moreover, as a hadron by itself becomes more rare, the 
 high-$p_T$  tail of the spectrum is  more strongly suppressed
 due to restrictions of the canonical ensemble.

Figure \ref{fig3} confirms our expectations on the applicability of the 3FD model. 
In peripheral collisions the total number of participants becomes smaller. 
Therefore, the applicability of the grand-canonical approximation 
and hence of the 3FD model worsens. 
It results in earlier (i.e. at lower $p_T$) overestimation of the tails of the $p_T$ spectra.
The fact that the 3FD model indeed fails beyond the range of its applicability 
gives us confidence in the 3FD results in the range of their  applicability.

\begin{figure}[thb]
%\vspace*{-3mm}
\includegraphics[width=8.6cm]{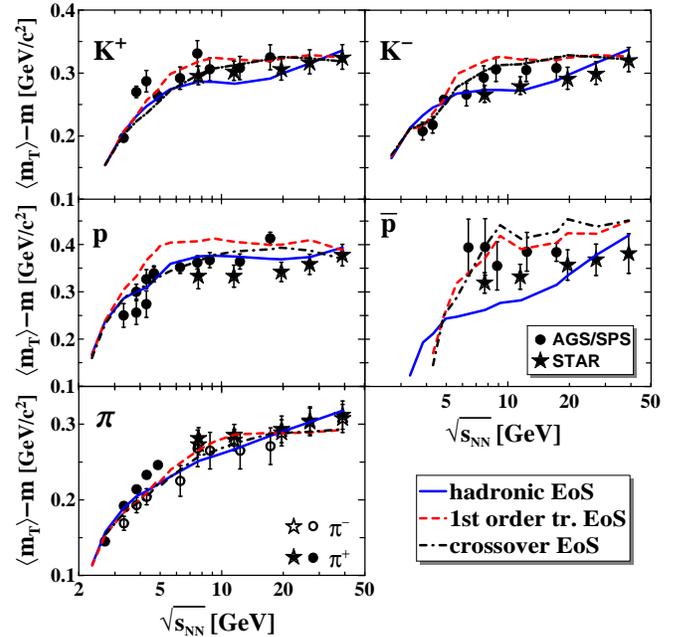}
%\vspace*{-49mm}
 \caption{(Color online)
Mean transverse mass (minus particle mass) at midrapidity for 
various particles 
produced in central ($b=2$ fm) Au+Au 
collisions 
%(at AGS and RHIC energies, $b=$ 2 fm) and Pb+Pb  (at SPS energies, $b=$ 2.4 fm) 
as functions of the collision energy.   
Experimental data are taken from Refs. 
\cite{Ahle:1999uy,Klay:2001tf,Barrette:1999ry,Ahle:1999in,Ahle:1998jc}
for AGS, 
\cite{Afanasiev:2002mx,Alt:2007aa,Alt:2006dk,Anticic:2004yj}
for SPS and  \cite{Adamczyk:2017iwn} for BES-RHIC energies.  
%\vspace*{-0mm}
} 
\label{fig4}
\end{figure}
\begin{figure*}[htb]
%\vspace*{-14mm}
\includegraphics[width=18.0cm]{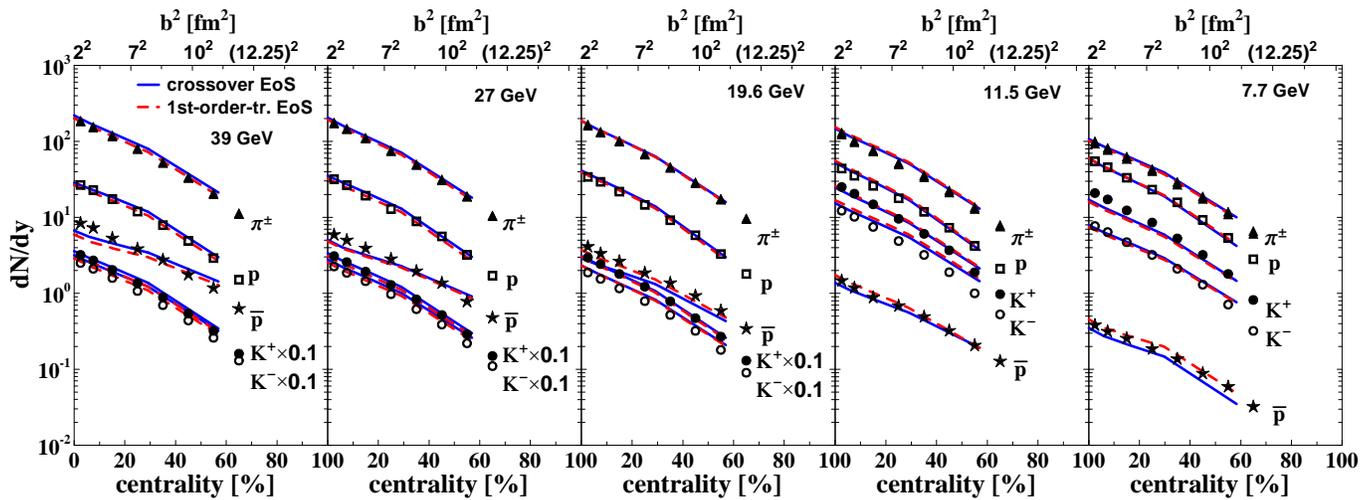}
 \caption{(Color online)
Midrapidity  densities $dN/dy$ of various  particles 
produced in  Au+Au collisions at $\sqrt{s_{NN}}=$ 7.7 -- 39 GeV
as functions of the centrality (and impact parameter squared, cf. upper scale)
 predicted  by 3FD model with two considered deconfinement EoS's. 
Experimental STAR data are from Ref. \cite{Adamczyk:2017iwn}.
}
\label{fig5}
\end{figure*}
\begin{figure}[htb]
%\vspace*{-14mm}
\includegraphics[width=6.2cm]{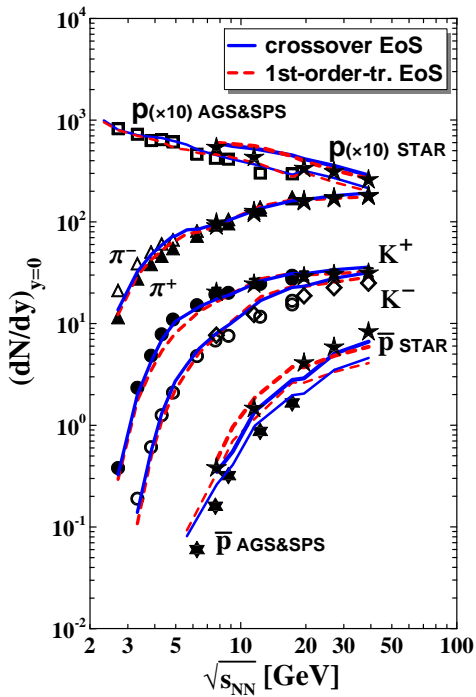}
 \caption{(Color online)
Midrapidity  densities $dN/dy$ of various particles produced  
 in central ($b=$ 2 fm) Au+Au collisions  predicted
 by 3FD calculations with  two considered deconfinement EoS's 
as functions of the center-of-mass energy. 
Thin lines for $p$ and $\bar{p}$ represent densities without contributions of weak decays 
(corresponding to NA49 data \cite{Compilation-NA49,Blume:2011sb}), 
while thick lines (for $p$ and $\bar{p}$) display densities 
including the weak-decay contributions which are relevant 
to the STAR data \cite{Adamczyk:2017iwn}.
Experimental data are from compilation of Ref. \cite{Andronic:2005yp}
complemented by recent data from STAR collaboration \cite{Adamczyk:2017iwn} (stars and open diamonds) 
and  latest update of the  
compilation of NA49 results \cite{Compilation-NA49,Blume:2011sb}.
}
\label{fig6}
\end{figure}
\begin{figure}[thb]
\includegraphics[width=6.5cm]{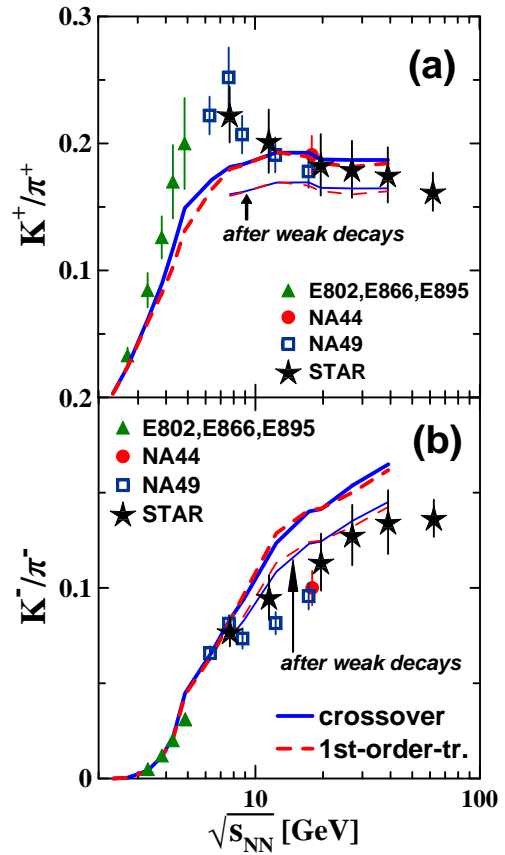}
 \caption{(Color online)
The energy dependence of ratios of the midrapidity densities $dN/dy$: 
(a) $K^+$ to $\pi^+$ and
(b) $K^-$ to $\pi^-$. 
Compilation of experimental data is from Ref. 
\cite{Blume:2011sb}  
complemented by recent STAR data \cite{Adamczyk:2017iwn} and \cite{Abelev:2008ab}
for $\sqrt{s_{NN}}=$ 62 GeV (stars).
Calculated ratios with pions including contribution from weak decays 
are also displayed by thin lines.   
} 
\label{fig_ratios}
\end{figure}

\subsection{Mean Transverse Mass}
\label{Mean Transverse Mass}

The collision-energy dependence of the mean transverse mass is 
an economical way to present the excitation function of the $p_T$ spectrum.  
Excitation functions of the average transverse masses 
($\langle m_T\rangle -m$) of  
protons, pions, antiprotons and kaons are presented in Fig. \ref{fig4}.  
The presented excitation functions are in fact predictions for the 
BES-STAR data because they were published in Ref. \cite{Ivanov:2013yla}
before the STAR data appeared. 
In Fig. \ref{fig4} we just compare the predictions of Ref. \cite{Ivanov:2013yla}
with the recent STAR data \cite{Adamczyk:2017iwn}. 
%Comparison with old AGS/SPS data
%\cite{Ahle:1999uy,Klay:2001tf,Barrette:1999ry,Ahle:1999in,Ahle:1998jc,Afanasiev:2002mx,Alt:2007aa,Alt:2006dk,Anticic:2004yj}
%is also presented. 
A detailed discussion of the calculated mean transverse masses 
is also presented in Ref. \cite{Ivanov:2013yla}.

As seen from Fig. \ref{fig4} the step-like structure is well reproduced within all scenarios, 
except for antiprotons within the hadronic scenario. This is a consequence the specific 
freeze-out implemented within the 3FD model \cite{Russkikh:2006aa}. 
The physical pattern behind this freeze-out resembles the
process of expansion of a compressed and heated classical
fluid into vacuum, mechanisms of which were studied both
experimentally and theoretically. The freeze-out is associated with evaporation from 
the surface of the expanding fluid and sometimes with  explosive 
transformation of the strongly superheated fluid to the gas. 
This freeze-out procedure results in a step-like excitation function of the 
effective freeze-out temperature \cite{Ivanov:2013yla}
which in its turn produces this step-like structure
in $\langle m_T\rangle$. 

Indeed, $\langle m_T\rangle$ can be very approximately presented as \cite{Mohanty:2003fy}
$$
\langle m_T\rangle - m \approx T_{\rm{eff}} + \frac{T_{\rm{eff}}^2}{m+T_{\rm{eff}}},
$$
where the inverse slope parameter $T_{\rm{eff}}$ 
can be very approximately related to the freeze-out temperature $T_{\rm{frz}}$ 
and transverse velocity $v_{\rm{tr}}$
$$
T_{\rm{eff}} \approx T_{\rm{frz}} + \frac{1}{2} m v_{\rm{tr}}^2. 
$$
The latter relation results from
the nonrelativistic limit of the blast-wave model \cite{Bondorf78,Siemens79,Schnedermann:1993ws}.
As seen, $\langle m_T\rangle$ indeed should follow the step-like behavior of $T_{\rm{frz}}$
if $v_{\rm{tr}}$ is approximately constant. This is actually the case for all particles and scenarios 
except for the antiprotons within the hadronic scenario. 
Antibaryons are dominantly produced from the baryon-free f-fluid. 
Because of long formation time of the f-fluid  within the hadronic scenario \cite{Ivanov:2013wha} 
(as compared with those within the deconfinement scenarios) 
the  transverse velocity $v_{\rm{tr}}$ in the f-fluid is not constant but steadily rises 
with the collision energy, therefore $\langle m_T\rangle$ rises too.

Figure \ref{fig4} demonstrates that
on average the crossover scenario still looks preferable, 
though the new STAR data are somewhat below those from SPS (for protons and antiprotons). 
It could seem that the crossover and hadronic scenarios reproduce (or fail to reproduce) 
the antiproton STAR data to the same extent. However, the slope of the crossover curve is 
definitely more adequate to the data. 
In fact, the overestimation of antiproton $\langle m_T\rangle$ is expected because of 
the overestimation high-$p_T$ tails of the $\bar{p}$ spectra discussed above, cf. Fig. \ref{fig2}. 
The first-order-transition scenario certainly fails for protons 
in the energy region where onset of deconfinement takes place in this EoS.

\subsection{Rapidity Densities}
\label{Rapidity Densities}

Midrapidity densities of various  particles 
produced in  Au+Au collisions at $\sqrt{s_{NN}}=$ 7.7 -- 39 GeV
as functions of the centrality %(and impact parameter squared, cf. upper scale)
predicted  by 3FD model are presented in Fig.  \ref{fig5}, where they are compared with 
the recent experimental STAR data \cite{Adamczyk:2017iwn}.
The results of calculations are in good overall agreement with the data except for 
kaons and antiprotons. Figure \ref{fig6} presents the midrapidity  densities
of these various  particles produced in  central ($b=$ 2 fm) Au+Au collisions but 
in a wider energy range $\sqrt{s_{NN}}=$ 2.7 -- 39 GeV.

The densities of positive kaons are underestimated at 7.7 GeV and mostly 
in central collisions, as already seen in $p_T$ spectra. 
This overestimation is not well resolved in Figs.  \ref{fig5} and \ref{fig6}, 
However, it is clearly visible in the $K^+/\pi^+$ ratio in Fig. \ref{fig_ratios}. 
The $K^-$ densities  are overestimated at $\sqrt{s_{NN}}\gsim$ 8 GeV, as seen 
from Fig. \ref{fig6} and more distinctly from $K^-/\pi^-$ ratio in Fig. \ref{fig_ratios}. 
As seen from Fig. \ref{fig2}, the soft parts of the $K^-$ $p_T$-spectra are reasonably well 
reproduced at all BES RHIC energies. 
The overestimation observed in Fig. \ref{fig6} 
may result from the overestimation of high-$p_T$ tails of $K^-$ spectra, cf. Fig. \ref{fig2}, 
which is the common problem for all rare probes described within the approximation of 
the grand-canonical ensemble. 
However, this explanation is not applicable at 39 GeV when negative and positive kaons 
are equally abundant and the  high-$p_T$ tail of the $K^-$ spectrum is quite well reproduced. 
Another possible source of this overestimation is extrapolation of 
the experimental $p_T$-spectra to a very low $p_T$-region in order 
to calculate the midrapidity density. If this experimental extrapolation differs from 
the 3FD prediction for the low-$p_T$ spectrum, the difference between the experimental and 
3FD midrapidity densities can be substantional.

Antiproton density is a delicate issue. 
Of course, the above discussed problems of the overestimation 
of high-$p_T$ tails and the extrapolation to the low-$p_T$ region take place for the antiprotons too. 
At lower collision energies the 3FD results overestimate 
the data (cf. Fig. \ref{fig6}), as expected within the approximation of 
the grand-canonical ensemble. 
However, at higher energies the data are already underestimated 
in the central collisions (Figs.  \ref{fig5} and \ref{fig6}) while they reasonably well agree with 
the 3FD results at larger centralities (impact parameters), see Fig.  \ref{fig5}.
This is already contrary to our expectations. Recently the production of anti-baryons was studied
within the Parton-Hadron-String Dynamics (PHSD) transport approach \cite{Seifert:2017oyb}. 
It was found that the PHSD results are also far from being perfect in reproduction of the 
antiproton data. This implies that we still do not well understand the mechanism of the $\bar{p}$ 
production.

The particle ratios, Fig. \ref{fig_ratios}, allow us to more distinctly see the problems 
in the description of the particle yields. 
The upper panel of Fig. \ref{fig_ratios} demonstrates the $K^+/\pi^+$ ratio with the 
well-known horn structure. 
Several statistical
models \cite{Andronic:2005yp,Andronic:2008gu,Bugaev:2013sfa} have succeeded in reproducing this
experimental trend, but they
can provide only a statistical description of the heavy-ion
collision process. There were no
conclusive interpretations of the "horn" from dynamical
approaches. 
The 3FD model is not an exception, 
it does not reproduce the horn anomaly in the 
$K^+/\pi^+$ ratio \cite{Ivanov:2013yqa}. 
Note that only deconfinement transitions are included 
in the EoS's tested within the 3FD model. 
Recently, it was suggested that the chiral phase 
transition within the PHSD model \cite{Palmese:2016rtq}
results in a semiquantitative  description of the horn anomaly. 
Restoration of the chiral symmetry enhances the strangeness production. 
It is equally important that this strangeness enhancement survives in the 
hadronic phase, as demonstrated in Refs. \cite{Palmese:2016rtq}. 
Another possible mechanism to explain the ``horn'' effect was developed in Ref. \cite{Dubinin:2016wvt}
where the horn anomaly was associated with the occurrence of an anomalous mode for mesons 
composed of quarks with unequal masses in the course of the Mott transition within the PNJL model. 
It is still unclear if the produced effect survives in the 
hadronic phase, i.e. till the freeze-out.

The lower panel of Fig. \ref{fig_ratios} demonstrates the $K^-/\pi^-$ ratio
which is overestimated within the 3FD model because of too high predicted 
$K^-$ yield. 
In order to demonstrate the effect of the weak decays
Figs. \ref{fig6} and \ref{fig_ratios} also display results with and without 
contributions of weak decays at BES-RHIC energies.

\begin{figure}[thb]
\includegraphics[width=6.cm]{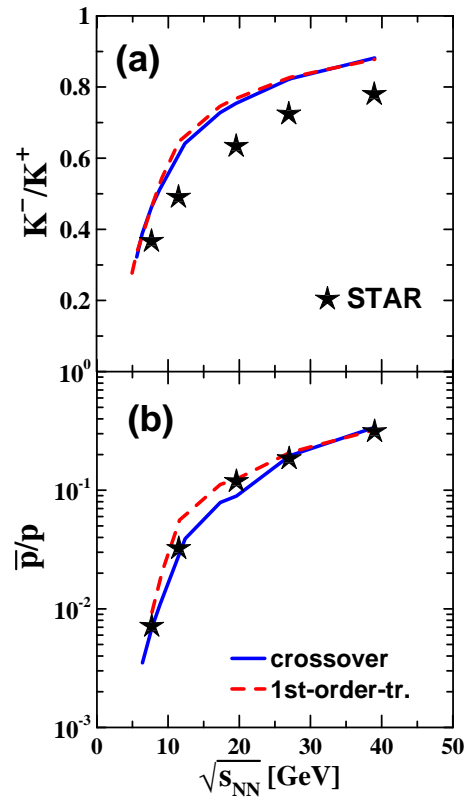}
 \caption{(Color online)
The energy dependence of ratios of the midrapidity densities $dN/dy$: 
(a) $K^-$ to $K^+$ and (b) $\bar{p}$ to $p$. 
Experimental STAR data are from Ref. \cite{Adamczyk:2017iwn}. 
} 
\label{fig_ratios-2}
\end{figure}

Figure \ref{fig_ratios-2} demonstrates comparison of the 3FD results on the $K^-/K^+$ and $\bar{p}/p$
ratios with the corresponding STAR data \cite{Adamczyk:2017iwn}. The calculated $K^-/K^+$ ratio
overestimates the data because of the above mentioned excess of $K^-$ at $\sqrt{s_{NN}}>$ 8 GeV. 
This is similar to that presented in panel (b) of Fig. \ref{fig_ratios}. The $K^-/K^+$ overestimation 
at $\sqrt{s_{NN}}=$ 7.7 GeV results from the underestimation of the midrapidity $K^+$ yield that 
is also clearly manifested in panel (a) of Fig. \ref{fig_ratios}. The reproduction of the $\bar{p}/p$
ratio [panel (b) of Fig. \ref{fig_ratios-2}] is quite satisfactory, which is partially the effect of 
the logarithmic scale of the figure.

\section{Summary}
\label{Summary}

Recent STAR data \cite{Adamczyk:2017iwn} on 
the bulk observables in the BES-RHIC
energy range were analyzed within the 3FD  model.
The simulations  
were performed with different 
equations of state: the purely hadronic EoS \cite{gasEOS}  
and two versions of the EoS involving the   deconfinement
 transition \cite{Toneev06}, i.e. the first-order phase transition
and the smooth crossover one. 
It was demonstrated that the purely hadronic EoS fails to reproduce 
the data. Therefore, we focused on the deconfinement EoS's. 
%which, however, do not involve the chiral phase transition   
%and critical point by construction. 

A good overall reproduction of the bulk observables of Ref. \cite{Adamczyk:2017iwn} 
is found within the deconfinement scenarios, of course, within the range of applicability 
of the 3FD model. Note that some of the discussed 3FD results are in fact predictions 
because they were published \cite{Ivanov:2013yqa,Ivanov:2013yla} before the appearance 
of the recent STAR data \cite{Adamczyk:2017iwn}. 
The crossover EoS tunes out to be slightly preferable. 
For this reproduction a quite strong friction in the QGP is required. 
This friction results in rather high initially equilibrated baryon densities in the central region
even at high collision energies.
However, this overall reproduction is not perfect. 
The main problem is the failure to reproduce the horn anomaly in the $K^+/\pi^+$ ratio. 
It is very probable that this anomaly is a signature of the chiral phase transition
\cite{Palmese:2016rtq,Dubinin:2016wvt}
which is absent in the present scenarios. 
It would be instructive to study the effect of the hadronic afterburner 
at the BES-RHIC energies within the THESEUS event generator \cite{Batyuk:2016qmb}.  
The afterburner may, at least partially, solve the problems with $K^-$ and $\bar{p}$
observables. It was found earlier \cite{Batyuk:2016qmb} that the hadronic afterburner only weakly affects the bulk 
observables of protons, pions and kaons, though at lower collision energies.

It should be noted that the 3FD model does not need two separate freeze-outs 
(i.e. kinetic and chemical ones) to describe 
the BES-RHIC data, as well as the older AGS and SPS data. 
A unified freeze-out is applied at all energies \cite{3FD,Russkikh:2006aa}. 
At the freeze-out stage the baryon-rich fluids 
are already unified while the net-baryon-free fluid still keeps its identity. Thus, 
the frozen-out system turns out to be kinetically equilibrated to a good extent, 
because the relative velocities of the baryon-rich and baryon-free fluids are small
\cite{Ivanov:2017xee}. At the same time the chemical equilibrium is absent: two overlapped
equilibrated fluids do not provide a chemical equilibrium in the composed system. 
This way the simultaneous description of $p_T$ spectra and hadron abundances is achieved.

\vspace*{3mm} {\bf Acknowledgments} \vspace*{2mm}

%Fruitful discussions with D.N. Voskresensky 
%are gratefully acknowledged. 
This work was carried out using computing resources of the federal collective usage center «Complex for simulation and data processing for mega-science facilities» at NRC "Kurchatov Institute", http://ckp.nrcki.ru/.
Y.B.I. was supported by the Russian Science
Foundation, Grant No. 17-12-01427.
A.A.S. was partially supported by  the Ministry of Education and Science of the Russian Federation within  
the Academic Excellence Project of 
the NRNU MEPhI under contract 
No. 02.A03.21.0005. %, 27.08.2013. 

\end{document}